\documentclass[twocolumn]{aa} 
\usepackage[dvips]{color}
\usepackage{graphicx}
\usepackage{txfonts}
\newcommand{\leoii}{\hbox{Leo\,II}}

\begin{document}
   \title{The   blue plume population in  dwarf spheroidal
   galaxies:}

   \subtitle{genuine blue stragglers or young stellar population?\thanks{Based
   on archival ESO and HST data.} }


\author{Y. Momany \inst{1}
\and
E.V. Held \inst{1}
\and
I. Saviane \inst{2}
\and
S. Zaggia \inst{1}
\and
L. Rizzi \inst{3}
\and
M. Gullieuszik \inst{1}
} 
\offprints{Y. Momany}
\institute{INAF: Osservatorio Astronomico di Padova, vicolo dell'Osservatorio 5, 35122 Padova, Italy \\
\email{yazan.almomany,enrico.held,marco.gullieuszik,simone.zaggia@oapd.inaf.it}
\and
European Southern Observatory, A. de Cordova 3107, Santiago, Chile \\
\email{isaviane@eso.org}
\and
Institute for Astronomy, 2680 Woodlawn Drive, Honolulu, HI 96822, USA \\
\email{rizzi@ifa.hawaii.edu}
}
\date{Received December 23, 2006; accepted  April 3, 2007}


\abstract{}
{Blue stragglers (BSS)  in   the  Milky Way field  and   globular/open
clusters are   thought to  be  the  product  of  either primordial  or
collisional binary  systems.    In the  context of   dwarf  spheroidal
galaxies it  is hard to  firmly disentangle  a genuine  BSS population
from young main sequence (MS) stars tracing a $\sim  1-2$ Gyr old star
forming episode.  }
%
%
{  Assuming that their blue plume populations are made of BSS, 
we  estimate the BSS  frequency ($F^{\rm BSS}_{\rm HB}$; as normalized
to   the  horizontal  branch star    counts)   for 8 Local  Group  non
star-forming  dwarf galaxies, using a compilation  of ground and space
based photometry.}
%
{ (i)   The  BSS frequency in   dwarf galaxies,  at  any given ${
M_{V}}$, is  always higher than  that in globular  clusters of similar
luminosities;  (ii) the BSS frequency  for the lowest luminosity dwarf
galaxies is in excellent agreement with that observed in the Milky Way
halo  and  open clusters; and    most  interestingly (iii)  derive   a
statistically    significant    $F^{\rm  BSS}_{\rm      HB}-{  M_{V}}$
anti-correlation for dwarf galaxies,    similar to that   observed  in
globular clusters.}
{The low density, almost   collision-less, environments of our   dwarf
galaxy  sample   allow  us  to  infer (i)  their  very   low dynamical
evolution; (ii) a   negligible  production  of collisional BSS;    and
consequently    (iii) that  their blue   plumes    are mainly made  of
primordial binaries.
The dwarf galaxies  $F^{\rm  BSS}_{\rm HB}-{ M_{V}}$  anti-correlation
can be used as a discriminator:  galaxies obeying the anti-correlation
are more likely  to possess genuine  primordial BSS rather than  young
main  sequence stars. 
}
{}


\maketitle
%

\section{Introduction}

In the context of  Galactic Globular clusters studies, blue stragglers
(BSS: a hotter and  bluer   extension of normal  main sequence  stars)
represent the highest manifestation  of the interplay  between stellar
evolution and stellar dynamics (Meylan \& Heggie \cite{meylan97}).
The origin of BSS is sought as either {\it primordial binaries} coeval
with  the cluster formation epoch,  or  to a continuous production (in
successive epochs)  of    {\it collisional  binaries}   due to
dynamical  collisions/encounters  experienced  by  single/binary stars
throughout the life of the cluster.
Ever  since their   identification in  (Sandage \cite{sandage53}), BSS
have been subject  of many photometric and  a handful of spectroscopic
(Ferraro   et    al.  \cite{ferraro06a},   and    references  therein)
studies. Nevertheless, BSS remain quite difficult to understand in the
light of {\em a single} comprehensive scenario.
Indeed, it is sometimes necessary to  invoke {\em both} the primordial
and collisional mechanisms to explain the BSS distribution in the same
cluster.
For example the bimodal distribution of BSS in 47Tuc (Ferraro et al.
\cite{ferraro04}; highly peaked in the  cluster center, rapidly
decreasing at intermediate radii,  and finally rising again at  larger
radii) can be interpreted as   evidence of two formation scenarios  at
work:
primordial binaries at the cluster periphery (where it is easier
 for them  to survive)  and collisional  binaries  at the center
(where it is easier to form).

A spectroscopic  survey by Preston    \& Sneden (\cite{preston00})  of
Milky  Way field  blue metal-poor  stars suggested  that over  60\% of
their sample  is made up  by binaries, and that at  least 50\%  of
their blue metal-poor sample are BSS.
Piotto et  al.  (\cite{piotto04}) presented a  homogeneous compilation
of $\sim3000$ BSS (based on HST  observations of 56 globular cluster),
and derived a significant and rather puzzling anti-correlation between
the  BSS specific frequency and  the cluster total absolute luminosity
(mass).
That  is  to say   that more  massive  clusters are  surprisingly  BSS
deficient, as if their higher collision rate had no correlation with
the production of collisional BSS.
Another puzzling observable is  that the  BSS  frequency in  Milky Way
(MW) field is  at least an order of magnitude larger than that of
globular clusters.
Recently De Marchi et al.  (\cite{demarchi06}) presented a photometric
compilation for  Galactic open  clusters with $-6\le{ M_{V}}\le-3$,
and   confirmed  an  extension  of  the   BSS  frequency-${ M_{V}}$
anti-correlation to the open clusters regime.
In an  attempt to explain these  observational trends, Davies  et
al.  (\cite{davies04}) envisage that while the  number of BSS produced
via  collisions tends  to  increase  with cluster mass,  becoming  the
dominant formation channel for clusters with ${\rm M_{V}\le-8.8}$, the
BSS number originating  from primordial binaries should  decrease with
increasing cluster mass.
Accounting for these two opposite  trends and binary evolution, Davies
et al.  (\cite{davies04}) models  are able to re-produce the  observed
BSS population,  whose  total {\it  number}  seems independent  of the
cluster mass.

Color-magnitude diagrams    (CMD) of typically   old dwarf  spheroidal
galaxies like Ursa  Minor (Feltzing et al.  \cite{feltzing99} and
Wyse  et  al.  \cite{wyse02}) show the   presence of a well-separated
blue plume of stars that very much resembles an old BSS population, as
that observed in globular and open clusters.
However,  in  the context of  dwarf galaxies  one {\em cannot} exclude
that blue plume stars  may include  genuinely young main sequence
(MS)  stars, i.e.  a   residual    star forming activity  (e.g.   Held
\cite{held05}, and references therein).
The BSS-young MS  ambiguity is  hard  to resolve,   and  has  been
discussed before  for Carina  (Hurley-Keller et  al. \cite{hurley98}),
Draco (Aparicio et al.   \cite{aparicio01}), and Ursa Minor (Carrera et
al. \cite{carrera02}).

In order  to investigate this ambiguity,  in this paper we measure the
BSS frequency in the dwarf  spheroidal galaxy \leoii\ and collect  BSS
counts in 8 other galaxies.
Dwarf spheroidals/irregulars in which there is {\em current} or recent
($\le500$ Myr) star formation are not considered.
For example, the  Fornax dwarf is  known to possess a young population
of $\sim200$ Myr (e.g. Saviane et al.  \cite{saviane00}) with young MS
stars {\em brighter} than the horizontal branch (HB) level.  Therefore
Fornax  cannot be considered  in  this paper.   The Sagittarius  dwarf
spheroidal, on the other hand, shows an extended blue plume, yet it does
not exceed the HB level, and is included in the study.
We also consider  the case of  the Carina dwarf and re-derive its
BSS frequency.
As discussed in Hurley-Keller et al.  (\cite{hurley98}) and Monelli et
al.   (\cite{monelli03}), Carina shows  evidence of  star formation in
recent epochs ($\sim1$ Gyr).
However, Carina     represents the first  case  in   which  the MS-BSS
ambiguity in dwarf galaxies  has been addressed:  a BSS  frequency was
derived and  compared with that in Galactic  globular clusters.  It is
also a  case in which  the ``youngest MS stars'' do  not exceed the HB
luminosity  level; i.e.  Carina  does not possess a significant recent
star formation rate.
As we shall see in Sect.~\ref{s_carina}, the Carina BSS frequency will
be most  useful for comparison  with other  dwarf  galaxies of similar
luminosities.

The BSS   frequency  is  therefore collected    for Sagittarius,
Sculptor,  Leo~II, Sextans, Ursa  Minor, Draco, Carina, Ursa Major and
Bo{\"o}tes and these  compared   internally, and  externally   (with that
reported for Galactic halo,  open and globular clusters).  This allows
us to address the dependence of the  BSS frequency on environment from
a wider perspective. 
%

\begin{figure}
\centering
\includegraphics[width=9.cm,height=10cm]{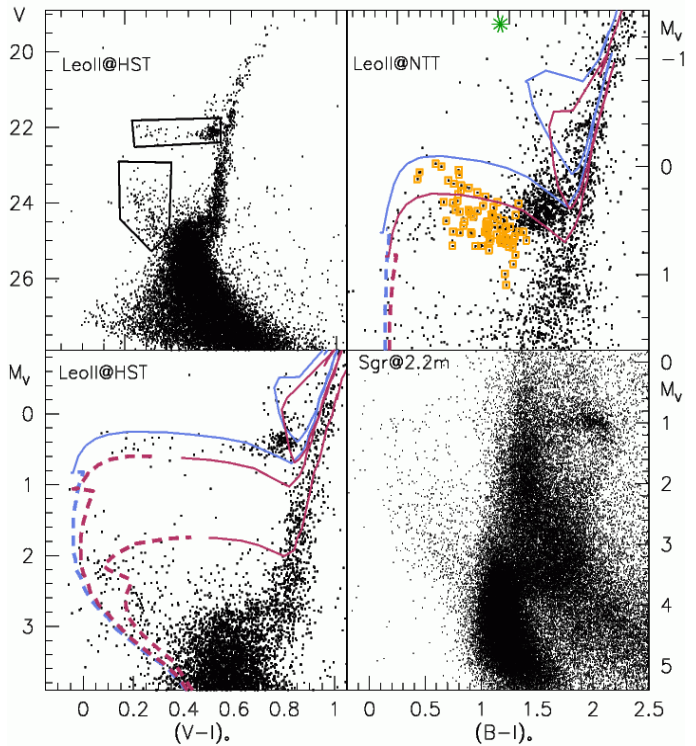}
\caption{Left panels display the HST CMDs of \hbox{Leo\,II},
upon which we  highlight the BSS and  HB selection boxes  and 2.5, 1.2
and      1 Gyr,  [Fe/H]$=-1.3$    isochrones   from   Girardi et   al.
(\cite{girardi02}).  Right upper panel displays the ESO/NTT diagram of
\hbox{Leo\,II} along with 1.0  and 0.8 Gyr isochrones highlighting the
extension of  the  vertical clump sequence  (dashed lines  mark the MS
phase    while continuous lines  track the   post-MS evolution).  Also
plotted   are    the   RR  Lyrae   stars    from  Siegel  \&  Majewski
(\cite{siegel00}) and one (asterisk) of  the 4 anomalous Cepheids (the
remaining 3 are  outside the NTT  field).  Lower right panel  displays
the    ESO/2.2m   CMD of     Sagittarius highlighting     the Galactic
contamination and the extension of its blue plume.}
\label{f_leoii}
\end{figure}
%

\section{BSS frequency data points}

The dwarf galaxies we   study in this paper    span a large   range of
distances ($\sim25$  Kpc    for  Sagittarius to  $\sim200$    Kpc  for
{\hbox{Leo\,II}).
This   basically precludes the    availability of one homogeneous  and
large-area imaging data-set reaching $1-2$ magnitudes below the old MS
turn-offs (see however the  recent HST/WFPC2 archival survey  of dwarf
galaxies by  Holtzman et al.  \cite{holtzman06}).  Thus estimating the
BSS frequency  for   dwarf galaxies,  unfortunately,  must rely   on a
compilation from various sources.
We  present new  reductions  of  archival  imaging  from  ESO/NTT  and
HST/WFPC2   for   {\hbox{Leo\,II},    and  ESO/2.2m  $BVI$  Pre-Flames
$1^{\circ}\times~1^{\circ}$         WFI         mosaic             for
Sagittarius\footnote{Excluding  the   inner  $14\farcm\times~14\farcm$
region around M54.}.  These  were reduced and calibrated following the
standard  recipes in  Held et al.  (\cite{held99})  and Momany et  al.
(\cite{momany01eis}, \cite{momany02sag} and \cite{momany05}).
For the  remaining dwarf galaxies we estimate  the BSS  frequency from
either public photometric catalogs (Sextans by Lee et al.\cite{lee03})
or published photometry kindly provided by  the authors (Ursa Minor by
Carrera   et      al.   \cite{carrera02},  Draco     by   Aparicio  et
al. \cite{aparicio01}, Sculptor by Rizzi et al.
\cite{rizzi03}, Ursa  Major by Willman   et al.  \cite{willman05}, 
Bo{\"o}tes  by Belokurov et al.  \cite{belokurov06} and Carina by
Monelli et al.  \cite{monelli03}).
All  the photometric catalogs  extend  to and  beyond the galaxy  half
light  radius; i.e.  we cover a  significant  fraction of the galaxies
and therefore the estimated  BSS frequency should  not be  affected by
specific spatial gradients, if present.    The only exception is  that
relative to Sagittarius.
With  a core radius  of $\sim3.7^{\circ}$, the estimated BSS frequency
of our $1^{\circ}$ square degree field refers to less than 3.5\% areal
coverage of  Sagittarius, or a  conservative  $\sim6$\% of the stellar
populations.
Nevertheless, our Sagittarius catalog is one among very few wide-field
available  catalogs of Sagittarius  that reach the BSS magnitude level
with an appropriate completeness level, and it is worthwhile to employ
it in this BSS analysis.

A delicate aspect of performing star counts is estimating the Galactic
foreground/background contribution in the covered area.  To compute it
in a homogeneous way we  made use of the  {\sc Trilegal} code (Girardi
et  al.  \cite{girardi05}) that  provides synthetic stellar photometry
of the Milky Way components (disk, halo, and bulge).
Star counts were performed on  the simulated diagrams (using the  same
selection boxes)  and subtracted  from the  observed  HB and BSS  star
counts.
We calculate the specific frequency of BSS  (normalizing the number of
BSS to the HB) as: $F_{\rm  HB}^{\rm BSS}={\rm \log}(N_{\rm BSS}/N_{\rm
HB})$.
We  remind  the  reader that  uncertainties   in  the (i)  photometric
incompleteness   correction, (ii)  foreground/background  subtraction,
(iii)   possible overlap  between   old and  intermediate age  stellar
population around the HB level,   and (iv) confusion between BSS   and
normal MS  stars, are all unavoidable problems  and affect the present
and similar studies.
The reported  error bars account   for the propagation  of the Poisson
errors on the  star counts, but  mostly reflect the dependence  on the
uncertainty in properly defining the HB and BSS selection boxes.
\subsection{Dwarf galaxies with a non-standard BSS population}
The  Leo~II dwarf might  be a   prototype of a galaxy whose blue
plume    properties differ from the   classical  BSS sequence found in
globular clusters.
Figure~\ref{f_leoii} displays the HST/NTT CMDs of {\hbox{Leo\,II} upon
which we highlight the  BSS and HB  selection boxes.  The Leo~II
BSS specific frequency  was   estimated from  the HST  diagram,  whose
coverage is  comparable with  the   galaxy half light  radius,   about
$\sim2$ arcmin.
The most  notable feature is the detection  of a vertical extension in
correspondence of the red HB  region.  Stars forming this sequence are
usually  called      vertical  clump  stars    (VC,  see    Gallart et
al. \cite{gallart05}).
These are helium-burning  stars of few hundred  Myr to $\sim1$ Gyr old
population whose progenitors are to be searched in the blue plume.
Indeed, the relatively large area covered by  the NTT shows a CMD with
a  well defined VC  sequence   that can be    matched by $\sim1$   Gyr
isochrones.
Since in the context of dwarf galaxies one cannot exclude the presence
of an extended star formation, the  detection of VC  stars (as is also
the   case for Draco,  see  Aparicio  et al. \cite{aparicio01})  would
suggest that the blue plume population may well hide a genuinely young
main sequence.
This  possibility will     be  further investigated    in   a detailed
reconstruction of the star formation history of {\hbox{Leo\,II} (Rizzi
et al in prep.).

A second diagnostic in support of a non-standard  BSS sequence lies in
the very extension of the {\hbox{Leo\,II}  blue plume.  The luminosity
function of blue  stragglers in globular clusters   has been found  to
increase from a luminosity cutoff at $M_{V}\sim1.9$ toward the ancient
MS turn-off at $M_{V}\sim 4.0$ (Sarajedini \& Da Costa
\cite{sarajedini91}, Fusi Pecci et al. \cite{fusi92}).
In the  case of  \leoii, the cutoff   luminosity should  correspond to
$V\simeq23.8$,   whereas    we  observe it   at    brighter magnitudes
($V\simeq23.0$ and  $V\simeq  22.7$  in   the  HST and  NTT   diagrams
respectively), i.e. the BSS sequence  in \leoii\ is more extended than
that in globulars.

A third diagnostic is  that relative to the ratio  of BSS to {\em
blue }  HB stars  ($7,000\le~T_{\rm eff}~\le~11,500$ K: the lower
temperature limit marks the blue  border  of the RR Lyrae  instability
strip while the upper   temperature limit signs the  horizontal branch
truncation  in   dwarf spheroidal galaxies,   as  noted in   Momany et
al. \cite{momany04}, \cite{momany02}).
This   diagnostic   has   been   used   by   Hurley-Keller  et
al. (\cite{hurley98}) in favor of a  genuine $\sim1$ Gyr MS population
in Carina, and similarly can be applied for Leo~II.
Basically,   for    Galactic  globular    clusters   Preston   et  al.
(\cite{preston94}) find a BSS to blue  HB ratio of $\sim0.6$.  This is
however  much  lower than what  we  estimate for  Leo~II:  using the 4
HST/WFPC2 catalogs we find a BSS to blue HB ratio of $\sim9.2$.
%
%
A fourth diagnostic is the BSS to VC ratio  that we estimate to remain
constant ($\sim8.1$) for each of the 4 WFPC2 chips, as well as for the
the entire  WFPC2    catalog.  This particular  observable   does  not
necessarily unveil  the    true  nature   of   the  blue    plume   in
{\hbox{Leo\,II}, it however confirms  a tight correlation  between the
BSS and VC populations.

Lastly,    we recall the  discovery  of   four   intermediate age
anomalous Cepheids in {\hbox{Leo\,II} (Siegel \& Majewski
\cite{siegel00}).  These are explained  as due to either extremely low
metallicity variables or to mass  transfer (and possibly  coalescence)
in a close binary systems.
However, as discussed by Siegel  \& Majewski (\cite{siegel00}), should
the  anomalous Cepheids in  {\hbox{Leo\,II}  be due  to mass  transfer
binaries, then  the  number of blue  stragglers (in  the HST field) is
predicted to be $\sim0.5-5$.
This is much  lower  than the  observed number  of BSS:  corrected for
incompleteness, we estimate a total of 175 BSS stars in the HST CMD.
Thus, accounting for: (i) the presence of VC stars; (ii) the extension
and (iii) number  of the  BSS  population in  {\hbox{Leo\,II} makes it
more  likely that  {\hbox{Leo\,II}  has experienced  an  extended star
formation history.
Similar conclusions   can  be applied   to  Sagittarius (possessing an
extended    blue plume whose  cutoff     luminosity is brighter   than
$M_V\sim1.9$,   as   illustrated in    Fig.~\ref{f_leoii})   and Draco
(possessing  VC    stars,     as  discussed   in    Aparicio  et
al. \cite{aparicio01}).

\begin{figure*}
\centering
\includegraphics[width=14cm,height=13cm]{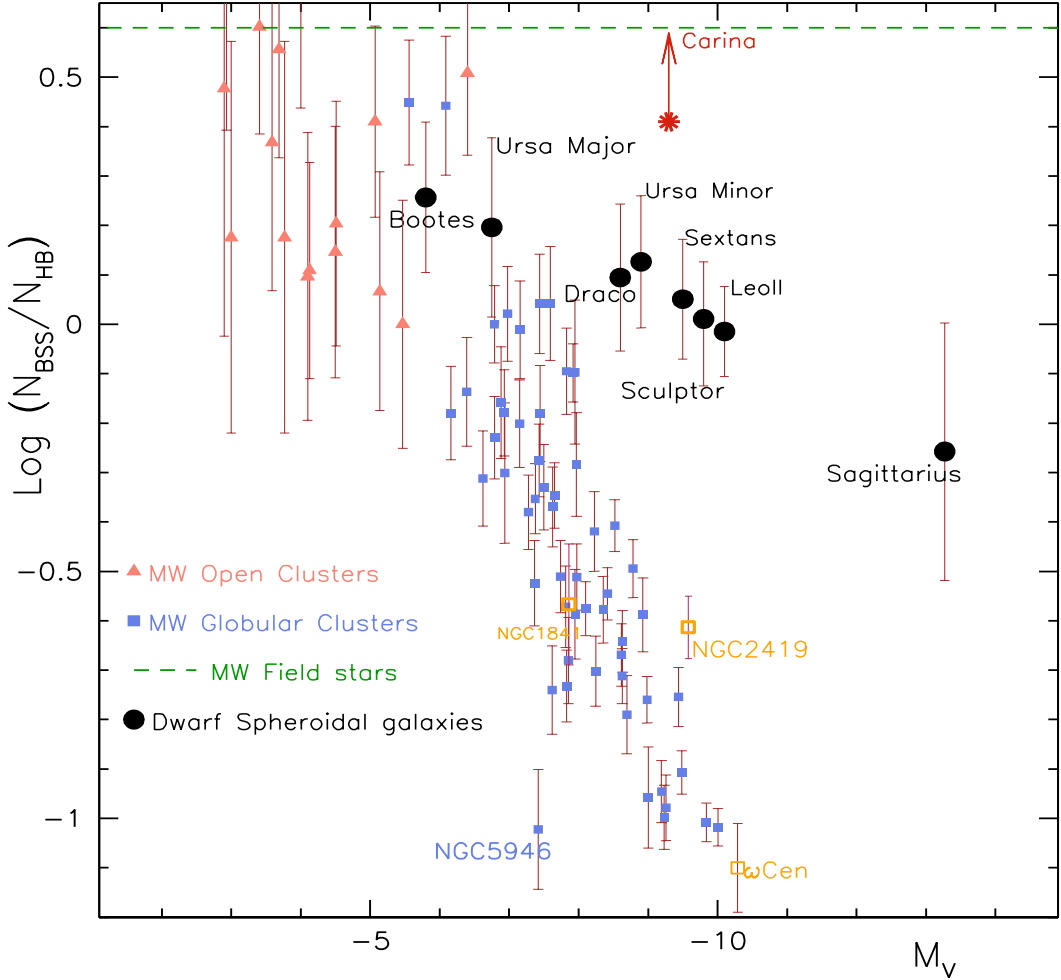}
\caption{The $F_{\rm BSS}$ vs ${ M_{V} }$ diagram for globular
clusters (Piotto et al.   \cite{piotto04}) open clusters (De Marchi et
al.  \cite{demarchi06}) and dwarf spheroidal galaxies.  The horizontal
line shows the  mean BSS frequency for Milky  Way field stars (Preston
\& Sneden \cite{preston00}). }
\label{f_fmv}
\end{figure*}

%
%
Despite all these arguments, it remains difficult to  rule out the BSS
interpretation.  
In this  regard, it is important  to note that the  strongest evidence
put   in favor of a recent   star formation episode in {\hbox{Leo\,II}
(i.e. the detection of VC stars) is a double-edge sword.
Indeed,  VC  stars   have  been detected    in  globular  clusters and
investigators needed  not to invoke a recent  star  formation in these
systems: the  presence  of   the  VC  sequence  could  and   has  been
interpreted as due to {\it evolved-BSS}.
As an example,  we  consider the case  of  M80.  Ferraro  et al.
(\cite{ferraro99}) derive a ratio of the BSS to evolved-BSS (or VC) of
$\sim7$.  This is very close to the BSS to VC ratio of $\sim8$ that we
estimate in Leo~II.
Mighell \& Rich (\cite{rich96}) were the  first to suggest that the VC
sequence in  {\hbox{Leo\,II}  can be due  to  the evolved BSS   in the
helium-burning phase.
Should this be the  case then one  need not to  explain the absence of
gas (Blitz \& Robishaw \cite{blitz00}) fueling a recent star formation
in {\hbox{Leo\,II}; since there should not be any.

\subsection{Dwarf galaxies showing a standard BSS population}
For few dwarf  galaxies there is no hint  for the presence of VC stars
(see the CMDs of Sextans by Lee et al.  \cite{lee03} and Ursa Minor by
Carrera et al.  \cite{carrera02}).
Although  foreground contamination might contribute  in veiling the VC
population, in general,  the absence of VC  or other intermediate age
indicators seemed to indicate a rather normal  BSS population in these
galaxies.

One  interesting  anomaly, however, has been   detected in the spatial
distribution of the BSS population in Sextans.
Lee et al.  (\cite{lee03}) find that brighter BSS  in Sextans are more
strongly concentrated  towards the  galaxy center,   while fainter BSS  are
lacking in  the central regions  and follow  the  distribution of  old
stars in the outer regions.
In the   context   of globular clusters    a  similar trend  is  often
attributed to the higher occurrence  of collisional binaries in higher
density  environments  (i.e.  the center) normally  producing brighter
and bluer BSS.
However,  the  collisional rate in  dwarf  spheroidals like Sextans is
much      lower   than  that        in    globular   clusters     (see
Sect.~\ref{s_discussion}),   and therefore   dynamical evolution    in
Sextans cannot account     for  a higher production of     collisional
binaries.   Thus, leaving aside  this particular distribution anomaly,
the overall blue plume    properties in galaxies like    Sextans, Ursa
Minor, Ursa Major and Bo{\"o}tes have been interpreted  as the old BSS
population.

\section{BSS frequency: analysis}
\label{s_discuss}

\subsection{BSS frequency in dwarf galaxies and globular clusters}
\label{s_carina}

We  now address the  BSS frequency for our  dwarf  galaxies sample and
make an internal comparison. For a wider  perspective, we also compare
the  overall  BSS frequency in dwarf  galaxies  with that  observed in
other stellar systems.
Figure~\ref{f_fmv}  displays the  $F_{\rm BSS}$  {\it vs}  ${ M_{V} }$
diagram for our  dwarf galaxy sample together  with the data-points of
Piotto    et     al.   (\cite{piotto04})    and   De   Marchi   et al.
(\cite{demarchi06}) for globular and open clusters, respectively.
Of the original  open cluster sample we only  plot clusters for  which
$\ge2$ BSS stars were found.
To the globular cluster sample we add the BSS frequency of $\omega$Cen
(as derived by Ferraro et al.
\cite{ferraro06}).   
The anomalies in   $\omega$Cen are multi-fold and
span a multiple MS, sub-giant and red giant branches (see Bedin et al.
\cite{bedin04}) and peculiar chemical enrichment (Piotto et al.
\cite{piotto05}), that are   often used  in  favor of  a  dwarf galaxy
origin.
Moreover we  estimate the BSS frequency  in  2  peculiar systems:
(i) NGC1841 (Saviane et al.  \cite{saviane03}) the LMC most metal-poor
and most distant ($\sim10$ Kpc from the LMC bar) globular cluster that
is also young  and incompatible with the LMC  halo  rotation; and (ii)
NGC2419 (Momany et  al.  in prep.)  a massive  MW cluster at  $\sim90$
Kpc from the Galactic center.  The three data-points are based on deep
HST/ACS, WFPC2 and ACS archival data, respectively.
Figure~\ref{f_fmv} clearly  shows  that, regardless of  their specific
peculiarities, $\omega$Cen,  NGC1841 and  NGC2419 are  consistent with
the general    globular  clusters $F^{\rm  BSS}_{\rm  HB}-{   M_{V} }$
anti-correlation.

Before turning  our  attention  to  the  BSS frequency  in dwarf
galaxies, we first  comment on the  case of NGC6717 and NGC6838  (the
two faintest globular clusters with the highest $F_{\rm HB}^{\rm BSS}$
frequency).
Located           at     ($l,b$)$=$($13^{\circ},-11^{\circ}$)      and
($50^{\circ},-5^{\circ}$), the two globular clusters can be subject to
significant bulge/disk contamination that was not accounted for in the
Piotto et al. analysis.
{\sc Trilegal} simulations  showed in fact  that a considerable number
of   Galactic  young MS stars  would   overlap  with the  clusters BSS
sequences and  this can  account for  their rather high  BSS frequency
with respect to globulars of similar ${ M_{V}}$.

Allowing  for   the  exclusion  of NGC6717  and   NGC6838,  it results
immediately that  the {\em lowest  luminosity dwarfs}  (Bo{\"o}tes and
Ursa Major)  would possess   a higher  $F_{\rm  HB}^{\rm  BSS  }$ than
globular clusters with similar ${ M_{V} }$.
Most interestingly, their  $F_{\rm  HB}^{\rm BSS }$  is  in fact fully
compatible with that observed in open clusters.
This compatibility   between   dwarf galaxies and   open  clusters may
suggest  that there exists a  ``saturation'' in the  BSS frequency (at
$0.3-0.4$) for  the lowest luminosity systems.  
Thus, the relatively  high $F_{\rm HB}^{\rm  BSS }$  of Bo{\"o}tes and
Ursa   Major adds    more   evidence in   favor  of  a    dwarf galaxy
classification of the 2 systems.
Indeed, although their   luminosities is  several times fainter   than
Draco or  Ursa   Minor,  the  physical  size   of  the  two   galaxies
($r_{1/2}\simeq220$ and  $250$  pc respectively)  exceeds that of more
massive galaxies like Ursa Minor ($r_{1/2}\simeq150$ pc).
%
%

Another interesting feature is  the significant difference between the
BSS frequency  of Carina with  that  derived for dwarf  galaxies  with
similar luminosity, i.e.   Draco,  Ursa Minor, Sextans, Sculptor   and
Leo~II.
Although it is only  a lower limit\footnote{It is  hard to account for
the  BSS    population originating   from the   older   and fainter MS
turn-off.}, the  ``BSS'' frequency  for  Carina  is  of great  help in
suggesting a threshold  near which a  galaxy BSS  frequency might hide
some level  of recent star formation.   The aforementioned  5 galaxies
however have a lower BSS frequency, a hint that these galaxies possess
a normal BSS population rather than a young MS.
This    confirms    previous   conclusions  for    Sextans   (Lee   et
al. \cite{lee03}) and Ursa  Minor (Carrera et  al.  \cite{carrera02}),
but is    in   contradiction    with   that of    Aparicio     et  al.
(\cite{aparicio01}) for Draco. However, the Aparicio et al. conclusion
was mainly based on the detection of the VC stars,  a feature that, as
we argued, remains  an ambiguous indicator.  Indeed,  Fig.~\ref{f_fmv}
shows that the BSS frequency  in Draco is very  close to that of  Ursa
Minor, a galaxy acceptably known to possess an old BSS population.

Lastly, leaving aside the extreme dynamical history of Sagittarius and
allowing  for    the  uncertainties   (due    to  the   heavy Galactic
contamination and the  relatively small sampled populations)  it turns
out  that  its blue  plume-HB frequency is  (i)  lower  than that of a
recently star-forming galaxy like Carina, and most interestingly; (ii)
in good agreement with the expected BSS  frequency as derived from the
$F^{\rm BSS}_{\rm HB}-{ M_{V}}$ anti-correlation  for the 7  remaining
galaxies in our sample.
Added to the  clear absence of  MS stars overlapping  or exceeding the
Sagittarius HB  luminosity level (see  Fig.~\ref{f_leoii}), we suggest
that  the Sagittarius blue  plume is a ``normal''  BSS sequence.  As a
matter of  fact, {\em Sagittarius is probably  the nearest system with
the  largest   BSS  population: over    2600 BSS  stars in  the  inner
$1^{\circ}\times~1^{\circ}$ field}.

To summarize, from Fig.~\ref{f_fmv}  one finds that  $F^{\rm BSS}_{\rm
HB}$ in  dwarf galaxies  is (i)  always higher than  that in  globular
clusters,  (ii) very close, for the  lowest luminosity dwarfs, to that
observed in the MW field and open  clusters, (iii) the Carina specific
$F^{\rm   BSS}_{\rm HB}$ frequency     probably sets a threshold   for
star-forming galaxies, and most interestingly, (iv)  shows a hint of a
$F^{\rm BSS}_{\rm HB}-{ M_{V}}$ anti-correlation.

\subsection{ A $F^{\rm  BSS}_{\rm   HB}-{ M_{V}}$ anti-correlation for
dwarf galaxies ?}   

We here explore the   statistical significance of a  possible  $F^{\rm
BSS}_{\rm   HB}-{\rm   M_{V}}$ correlation.    The  linear-correlation
coefficient   (Bevington  \cite{bevington69})   for  the   8  galaxies
(excluding Carina) data-points is $0.984$.
The corresponding probability that any  random sample of uncorrelated
experimental  data-points   would yield a correlation   coefficient of
$0.984$ is $<10^{-6}$.
Given the  greater uncertainties associated   with the Sagittarius BSS
frequency,  one  may  be  interested  in the   correlation coefficient
excluding the Sagittarius data-point.
In this case,   the resulting correlation coefficient  remains however
quite   high   ($0.972$) and  the probability     that the 7 remaining
data-points would randomly correlate is as low as $10^{-4}$.
Thus,  the  statistical significance of  the   $F^{\rm BSS}_{\rm HB}-{
M_{V}}$  anti-correlation in non  star-forming dwarf galaxies is quite
high.
We    follow  the   methods      outlined   in Feigelson   \&     Babu
(\cite{feigelson92}) and  fit least-squares  linear regressions.
In particular, the  intercept and slope regression coefficients  were
estimated through 5 linear models  (see the code  of Feigelson \& Babu
for         details)     the       average       of     which    gives
($a,b$)$=$($0.699\pm~0.081,0.070\pm~0.010$)                        and
($a,b$)$=$($0.631\pm~0.120,0.062\pm~0.014$)   including and  excluding
the Sagittarius data-point, respectively.
The  reported errors were  estimated  through {\sc Bootstrap} and {\sc
Jacknife} simulations so  as  to provide more   realistic $a$ and  $b$
errors.

However,   to firmly establish  this   $F^{\rm BSS}_{\rm HB}-{ M_{V}}$
anti-correlation one needs  to increase the  dwarf galaxies sample, in
particular at the two luminosity extremes.
Unfortunately there are not many non  star-forming dwarf galaxies with
$-13.3\le~M_{V}\le~-10.1$  (c.f.   table  14 in Mateo \cite{mateo98}),
and few exceptions may come from deeper imaging of galaxies like And~I
and And~II.
On    the  other hand, more   Local  Group  dwarf  galaxies  are being
discovered  in   the low luminosity  regime ($-8.0\le~M_{V}\le~-5.0$).
Deeper imaging of recently discovered galaxies  like Com, CVn~IIm, Her
and  Leo~IV   (Belokurov et  al.    \cite{belokurov07}) and  Willman~1
(Willman et al.  \cite{willman06})  are  needed to estimate  their BSS
frequency.
The importance of these  low-luminosity galaxies is easily  understood
once we exclude  Bo{\"o}tes and  Ursa Major   from the  BSS  frequency
correlation analysis.
In this case,  the   correlation  coefficient  for the   5   remaining
data-points is found to drop to $0.901$ having a probability of random
correlation as high as $1.5\times10^{-1}$.  Thus,  a final word on the
$F^{\rm BSS}_{\rm HB}-{ M_{V}}$  anti-correlation must await for  more
data-points at both luminosity extremes.  
%
%
%
%

\begin{figure}
\centering
\includegraphics[width=9cm,height=10cm]{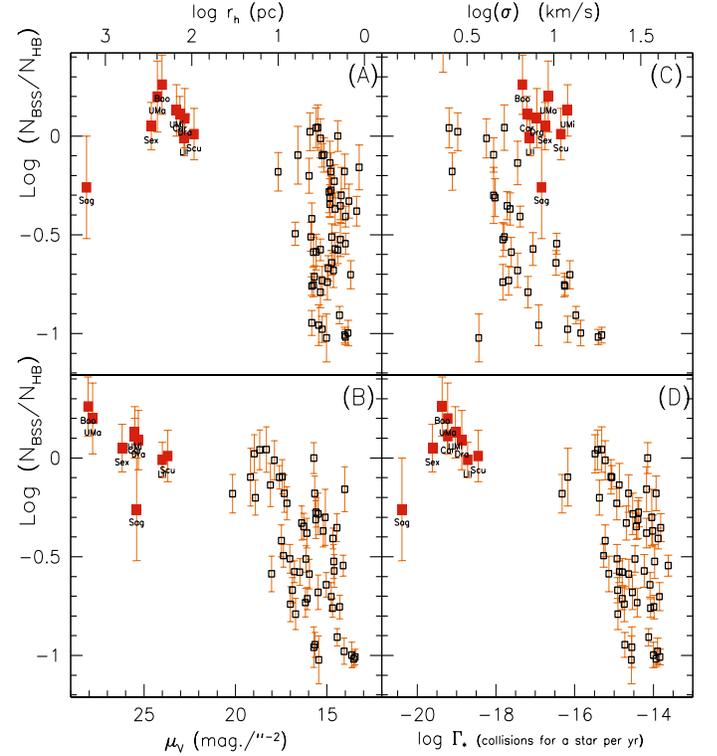}
\caption{The BSS  frequency as a function  of the 
half light radius (panel  { \it a}),   the central surface  brightness
(panel {\it b}),  the velocity dispersion (panel  {\it c}) and
the stellar  collision   factor (panel  {\it   d}).  See  text for
details.}
\label{f_extra}
\end{figure}
%

\section{Discussion and Conclusions}
\label{s_discussion}

For a sample of 8 non star-forming dwarf galaxies,  we have tested the
hypothesis that the blue plume  populations are made  of a genuine BSS
population (as   that observed in   open and   globular clusters)  and
estimated their frequency with respect to HB stars.
Should this assumption be incorrect (and  the blue plume population is
made of young MS stars) then one  would not expect an anti-correlation
between the   galaxies  total luminosity   (mass) and  the  blue plume
frequency, but rather a correlation between the two.
Instead, and  within the limits of  this and similar analysis, {\em we
detect a statistically  significant  anti-correlation between  $F^{\rm
BSS}_{\rm HB}$ and $M_{V}$}.
A  similar anti-correlation  has been  reported  for Galactic open and
globular clusters.  

A positive detection of vertical clump stars  {\em does not} provide a
clear-cut evidence in  favor of a recent  star formation episode in  a
dwarf galaxy. This is because the vertical clump population has been
detected in globular clusters and one {\em cannot} exclude that these
are evolved-BSS.
Thus, the main difference between  the blue plume (in non star-forming
dwarf galaxies)  and the BSS  sequence (in globulars clusters) is that
regarding their number.
Should  a dwarf galaxy  ``obey''  the $F^{\rm BSS}_{\rm  HB}-{ M_{V}}$
anti-correlation   displayed by our     sample then  its blue    plume
population is probably made of blue stragglers.

{\it Do dwarf galaxies  harbor a significant population of collisional
binaries ?}    The   answer is no.    This   relies on  the  intrinsic
properties of dwarf galaxies and  the consequent difference with those
of globular clusters. Indeed, it is enough to  recall that the central
luminosity   density   of   a   dwarf   galaxy    (e.g.  Ursa   Minor:
$0.006~L_{\odot}~pc^{-3}$   at  $M_V=-8.9$)   is  several  orders   of
magnitudes lower  than that found in a  typical globular cluster (e.g.
NGC7089  $\sim8000~L_{\odot}~pc^{-3}$  at $M_V=-9.0$).   This  implies
that  the  collisional parameter  of  dwarf galaxies  is very low, and
unambiguously point to the  much  slower dynamical evolution  of dwarf
galaxies.
To further emphasize this last point, we  search for $F^{\rm BSS}_{\rm
HB}$ dependencies on other dwarf galaxies structural parameters.
Figure~\ref{f_extra} plots the BSS frequency as a function of the half
light  radius  (panel {\it  a}),  the central  surface brightness
(panel {\it  b}), the velocity dispersion (panel  {\it c}) and
the stellar collision factor (panel {\it d})\footnote{For globular
clusters we make use of the Trager et al.  (\cite{trager95}) and Pryor
\&  Meylan (\cite{pryor93}) tables,   whereas for dwarf spheroidals we
use   the tables from    Mateo  (\cite{mateo98}) and  updated velocity
dispersions from recent   measurements   (Sculptor: Tolstoy et     al.
\cite{tolstoy04}, Sextans: Walker et al.
\cite{walker06}), Carina:    Koch et al.    \cite{koch07}, Bo{\"o}tes:
Mu{\~n}oz et al.  \cite{munoz06}, Ursa Minor: Mu{\~n}oz et al.
\cite{munoz05},  Ursa    Major:   Kleyna  et   al.    \cite{kleyna05},
Sagittarius: Zaggia et al.     \cite{zaggia04}, Draco: Mu{\~n}oz    et
al. \cite{munoz05}; Carina: Mu{\~n}oz et al. \cite{munoz06b}).}. 
Panels  {\it a} to {\it c} plot  the $F^{\rm  BSS}_{\rm  HB}$ as a
function of {\em observed} globular/dwarf galaxies quantities.
Panel {\it  a} shows that globular  clusters and dwarf spheroidals
form 2 quite distinct families.
This is  further confirmed in panel   {\it b}, although the central surface
brightness  distribution might  suggest   a  $F^{\rm   BSS}_{\rm  HB}$
connection of the two.
Panel {\it c}  shows a correlation between  $F^{\rm BSS}_{\rm HB}$ and the
central velocity dispersion  for globular clusters. This reflects  the
known globular  cluster   fundamental plane  relations, as  shown   by
Djorgovski (\cite{dj95}).  
Despite the similarities  in their velocity dispersion, dwarf galaxies
form a separate  group from globular  clusters, showing systematically
higher $F^{\rm BSS}_{\rm HB}$.

In panel  {\it d}  we  show $F^{\rm BSS}_{\rm HB}$  as a function of a  {\em
calculated} quantity:  the stellar specific collision parameter (${\rm
log}~\Gamma_{\star}$: the number of collisions per star per year).
More specifically, following Piotto et al.   (2004), for both globular
clusters   and dwarf galaxies  we  estimate ${\rm log}~\Gamma_{\star}$
from the central surface density and the system core size.
The   mean collisional   parameter  of   the  9  studied   galaxies is
$\simeq-19$.  The  lowest value  is  that  for Sagittarius with  ${\rm
log}~\Gamma_{\star}\simeq-20.2$, and this is due  to its very extended
galaxy core.
Compared  with the  mean value of  $-14.8$ for   the globular clusters
sample (see also  the lower panel  of Fig.~1 in Piotto  et al.  2004),
the estimated  number  of  collisions per  star  per  year in a  dwarf
spheroidal is $10^{-5}$ times lower.
%
%
{\em This almost  precludes the occurrence  of collisional binaries in
dwarf galaxies, and  one may  conclude  that genuine  BSS sequences in
dwarf galaxies are mainly made of primordial binaries}.

Not all primordial binaries, now present  in a dwarf galaxy, turn into
or are already in the form of BSS.
In particular, it  is   the low  exchange encounter probabilities   in
environments like the Galactic  halo or dwarf galaxies that guarantees
a {\it friendly}   environment and a slower consumption/evolution   of
primordial binary systems.
The BSS  production (via evolution off the  MS of the primary  and the
consequent mass-transfer to  the secondary that may  become  a BSS) is
still taking place in the present epoch and this  can explain the high
frequency of primordial BSS in dwarf  galaxies as well as the Galactic
halo.

%
%
%
%

%
Lastly, it  is  interesting  to note   how  the BSS frequency  in  the
low-luminosity  dwarfs      and open  clusters    (${\rm  \log}(N_{\rm
BSS}/N_{\rm  HB})\sim0.3-0.4$) is very close  to  that derived for the
Galactic halo (${\rm \log}(N_{\rm BSS}/N_{\rm HB})\sim0.6$) by Preston
\& Sneden.
The latter   value however  has  been derived  relying on  a composite
sample of only  62 blue metal-poor  stars that are  (i) distributed at
different line  of  sights;  (ii)  at  different distances; and   most
importantly,  (iii)  for which  no observational   BSS-HB star-by-star
correspondence can be established.
Thus, allowing for all these  uncertainties in the field BSS frequency
(see also  the discussion in  Ferraro et al. \cite{ferraro06}),  it is
safe to conclude that  the {\em observed open  clusters-dwarf galaxies
BSS frequency sets  a realistic, and observational  upper limit to the
primordial BSS frequency in stellar systems}.
}

\begin{acknowledgements}
We thank Alvio Renzini and Giampaolo Piotto for useful discussions that
helped  improve this paper.    We are also  grateful to  Belokurov V.,
Willman  B., Carrera R., Monelli M. and Aparicio  A. for kindly providing us their
photometric catalogs.

\end{acknowledgements} 

{}
\end{document}